\documentclass[preprint,12pt]{elsarticle}

\usepackage{graphicx}

\usepackage{amssymb}

\begin{document}

\begin{frontmatter}



\title{Nuclear activity and star formation properties of \\ Seyfert 2 galaxies}


\author{Rafanelli P.}
\author{D'Abrusco R.}
\author{Ciroi S.}
\author{Cracco V.}
\author{Di Mille F.}
\author{Vaona L.}
\address{Department of Astronomy, University of Padua, Vicolo dell'Osservatorio 3, Padua - Italy}

\begin{abstract}

In order to characterize the amount of recent or ongoing stellar formation in the circumnuclear region of active galaxies on a statistically sound basis, we have studied the stellar component of the 
nuclear spectra in three different samples of galaxies, namely Seyfert 2 galaxies (hereafter S2G), star-forming galaxies (SFG) and passive normal galaxies (NG),  i.e., no emission lines observed,  
using  Sloan Digital Sky Survey data (SDSS) \cite{sdss}. The stellar component of the observed spectra has been extracted using STARLIGHT \cite{cidfernandeS2G004}, which fits an observed 
spectrum with a model (template) spectrum obtained by combining a library of pre-defined simple stellar populations spectra, with distinct ages and metallicities. The resulting template spectra
for the different samples of galaxies have been compared to determine the features of the stellar emission component and to evaluate the presence and intensity of the star formation in the nuclear 
regions of different families of galaxies. From a first qualitative analysis it results that the  shape of the Spectral Energy Distribution (SED) of S2G and NG is  very similar, while that of SFG  is characterized  
by a strong blue excess. The presence  of  the 4000 \AA \ \ break in the spectra of S2G and NG  together with the lack of a strong blue continuum
clearly indicate the absence of ongoing star formation in the circumnuclear regions of S2G and obviously of NG. Anyway traces of a recent star formation history are evident in the spectra of S2G galaxies, which show a  4000 \AA \ \ break  systematically shallower than in NG.  \end{abstract}

\begin{keyword}

environment of AGN, stellar formation and spectral continuum distribution



\end{keyword}

\end{frontmatter}




\section{Selection of the samples}
\label{data}

The three samples of galaxies used in this work have been extracted from the SDSS DR7 spectroscopic dataset and classified on the basis of a set of classical spectroscopic 
diagnostic diagrams which exploit the flux ratios of spectral emission lines to determine the presence of AGN activity in the spectrum of the nuclear region of the galaxies. 
These samples have all been selected from a larger dataset composed of all galaxies observed spectroscopically and with redshift comprised in the 
range [0.04, 0.08], with signal-to-noise $S/N >  5$ in the spectral continuum at $\lambda = 5500 \AA$ with the additional requirement of a signal-to-noise ratio $S/N > 3$ 
in the strong emission lines $\mbox{[OIII]}\lambda 5007$, $\mbox{[OI]}\lambda 6300$, $\mbox{[OII]}\lambda 3727$, $\mbox{H}\beta$, $\mbox{H}\alpha$, $\mbox{[NII]}\lambda 6584$, 
$\mbox{[SII]}\lambda\lambda 6717\ 6731$, when observed. The selection in redshift has been performed in order to minimize the effect on the spectra of the fixed aperture of 
spectroscopic fibers used for SDSS observations (see \cite{kewley2006} for comparison) and allow the smallest possible contamination by stellar light emitted in the regions 
of the galaxies far outside from the nuclues. The first step of the selection process has been the separation of the sources without emission lines from the emission line galaxies. 
The 2000 galaxies with highest S/N ratios of this class have been chosen as members of the passive galaxies sample. The remaining emission line galaxies have been 
split in two subsamples, the first corresponding to the sources showing the signs of the presence of an AGN and the second containing SFG, Liners and composite galaxies by  
applying the diagnostic diagram based on the Oxygen lines flux ratios \citep{vaona2009}, using the fluxes measured by the SDSS and the empirical relation in the
$\mbox{[OI]}\lambda 6300/\mbox{[OIII]}\lambda 5007$ vs $\mbox{[OII]}\lambda 3727/\mbox{[OIII]}\lambda 5007$ diagram derived by \citet{vaona2009}. The spectra of the galaxies 
selected as AGN-hosting sources in the previous step have been retrieved by the spectroscopic SDSS database and corrected for galactic reddening using the values
of the extinction provided by the Nasa Extragalactic Database (NED) web-service\footnote{At the website {\it http://irsa.ipac.caltech.edu/applications/DUST/}}, based on the 
maps and tools discussed in \cite{schlegel1998}. Then, they have been reduced to the rest-frame using the value of the redshift as measured by the SDSS spectroscopic pipeline. The stellar continuum of the spectra of these presumptive Seyfert galaxies has been evaluated
using STARLIGHT (see below for details on this code), and subtracted from the observed spectra in order to retain the emission and absorption features of the spectra.
Afterthen, the spectral parameters of all emission lines with $S/N > 5$ have been re-measured using an original method performing a multi-gaussian fit of the the emission lines 
developed by L. Vaona during his PhD thesis \citep{vaona2009}. The galaxies showing in their spectra  $\mbox{H}\alpha$ and $\mbox{H}\beta$ line profiles broader than the $\mbox{[OIII]}\lambda 5007$ profile
have been rejected, and the remaining sample
has been classified using the classical spectroscopic diagnostic diagrams known as Veilleux-Osterbrock diagrams \citep{veilleux}. The last step of the selection process of the S2G 
galaxies has been the extraction of the galaxies which satisfy the empirical relations in the  $\mbox{[NII]}\lambda 6584/\mbox{H}\alpha$ vs $\mbox{[OIII]}\lambda 5007/\mbox{H}\beta$, $\mbox{[SII]}\lambda 6717/\mbox{H}\alpha$ vs $\mbox{[OIII]}\lambda 5007/\mbox{H}\beta$ and $\mbox{[OI]}\lambda 6300/\mbox{H}\alpha$ 
vs $\mbox{[OIII]}\lambda 5007/\mbox{H}\beta$ diagrams as determined by \citet{kewley2006} for a sample of galaxies observed spectroscopically in the SDSS DR4:

\begin{eqnarray}
\frac{0.61}{[\log \frac{\mbox{[NII]}}{\mbox{H}\alpha} - 0.47]} + 1.19 < \log \frac{\mbox{[OIII]}}{\mbox{H}\beta}\\
\frac{0.72}{[\log \frac{\mbox{[SII]}}{\mbox{H}\alpha} - 0.32]} + 1.30 < \log \frac{\mbox{[OIII]}}{\mbox{H}\beta}\\
\frac{0.73}{[\log \frac{\mbox{[OI]}}{\mbox{H}\alpha} - 0.59]} + 1.33 < \log \frac{\mbox{[OIII]}}{\mbox{H}\beta}
\end{eqnarray}

The SFG galaxies have been extracted from the original sample using the Oxygen diagram and then applying again the appropriate relations
in the Veilleux-Osterbrock diagnostic diagrams to the line fluxes as measured by the SDSS pipeline. To summarize, the final numbers of members 
of the NG, SFG and S2G galaxy samples used for the analysis described in section \ref{spectra} are 2000, 1302 and 1996 respectively. 

\section{Analysis of the spectra}
\label{spectra}

The spectra of the galaxies belonging to each of the three samples described in the previous section have been corrected 
for galactic reddening using the extinctions provided by the NED service and, similarly at what has been done for the selection of the S2G galaxies, 
reduced to the rest frame using the spectroscopic redshift measured by the SDSS. Then, all spectra have been realigned to match the dispersion of 
1 \AA/pixel in the spectral range $[3600\AA, 9200 \AA]$. At this point, STARLIGHT has been used to produce the template spectra accounting for 
the purely stellar emission in the nuclear regions of the galaxies. This code performs spectral synthesis on observed spectra, consisting in the 
decomposition of the spectrum in terms of a convenient superposition of a base of simple stellar populations of various ages and metallicities. The collection of 
stellar populations employed to perform the fits of the input spectra in this work consists in a set of 92 stellar populations extracted 
from the Bruzual-Charlot library of evolutionary synthesis models described \cite{bruzual2003} with 23 distinct ages spanning the interval $[10^{6}, 1.3\  10^{10}]$ 
years and 4 distinct metallicity values (Z = $4\  10^{-3}, 8\  10^{-3}, 2\  10^{-2}, 5\  10^{-2}$), as shown in table 1.

\begin{table}
\label{table1}     
\centering                 
\begin{tabular}{c c c c c c c c c}        
\hline\hline                
 & Age (yrs) & $Z$ &  & Age (yrs) & $Z$ &  & Age (yrs) & $Z$\\    
\hline                      
 1   &    $1.0\ 10^{6}$         & $4\  10^{-3}$ 	& 32    &  $4.0 \  10^{7}$     &  $8\  10^{-3}$ &  63   &  $1.4\  10^{9}$          & $2\  10^{-2}$ \\
 2    &   $3.2\  10^{6}$   	& $4\  10^{-3}$ 	& 33    &  $5.5 \  10^{7}$     &  $8\  10^{-3}$ & 64    &  $2.5\  10^{9}$          & $2\  10^{-2}$\\
 3    &   $5.0\  10^{6}$   	& $4\  10^{-3}$ 	&  34   &  $1.0 \  10^{7}$  	  &  $8\  10^{-3}$ & 65   &  $4.3\  10^{9}$           &  $2\  10^{-2}$ \\
 4    &   $6.6\  10^{6}$   	& $4\  10^{-3}$ 	& 35    &  $1.6 \  10^{8}$     &  $8\  10^{-3}$ & 66    &  $6.3\  10^{9}$          &  $2\  10^{-2}$ \\
 5    &   $8.7\  10^{6}$  	& $4\  10^{-3}$ & 36    &  $2.9\  10^{8}$       &  $8\  10^{-3}$ &  67  &  $7.5\  10^{9}$           &  $2\  10^{-2}$\\
 6     &  $1.0\  10^{7}$         & $4\  10^{-3}$ & 37   &  $5.2\  10^{8}$       &  $8\  10^{-3}$ & 68   &  $1.0\ 10^{10}$         &  $2\  10^{-2}$\\
 7     &  $1.5 \  10^{7}$       & $4\  10^{-3}$	& 38   &  $ 9.0\  10^{8}$       &  $8\  10^{-3}$ & 69   &  $1.3\  10^{10}$         &  $2\  10^{-2}$\\
 8     &  $2.5 \  10^{7}$       & $4\  10^{-3}$ 	& 39   &  $1.3\  10^{8}$        &  $8\  10^{-3}$ & 70   &  $1.0\ 10^{6}$            &  $5\  10^{-2}$ \\
 9     &  $4.0 \  10^{7}$       & $4\  10^{-3}$ & 40   &  $1.4\  10^{9}$       &  $8\  10^{-3}$ & 71    &  $3.2\  10^{6}$           &  $5\  10^{-2}$\\
 10   &  $5.5 \  10^{7}$       & $4\  10^{-3}$ & 41   &  $2.5\  10^{9}$        &  $8\  10^{-3}$ & 72   &  $5.0\  10^{6}$          & $5\  10^{-2}$\\
 11   &  $1.1 \  10^{7}$  	& $4\  10^{-3}$  & 42   &  $4.3\  10^{9}$        &  $8\  10^{-3}$ & 73   &  $6.6\  10^{6}$         & $5\  10^{-2}$\\
 12   &  $1.6 \  10^{8}$   	&  $4\  10^{-3}$ & 43   &  $6.3\  10^{9}$        & $8\  10^{-3}$ & 74    &  $8.7\  10^{6}$         &  $5\  10^{-2}$\\
 13   &  $2.9\  10^{8}$  	&  $4\  10^{-3}$ & 44   &  $7.5\  10^{9}$         &  $8\  10^{-3}$  &75  &  $1.0\ 10^{7}$          & $5\  10^{-2}$\\
 14   &  $5.1\  10^{8}$   	&  $4\  10^{-3}$ & 45   &  $1.0\ 10^{10}$        &   $8\  10^{-3}$ & 76 &  $1.5\  10^{7}$         &  $5\  10^{-2}$ \\
 15   &  $ 9.0\  10^{8}$  	&  $4\  10^{-3}$ & 46   &  $1.3\  10^{10}$       &  $8\  10^{-3}$ & 77  &  $2.5 \  10^{7}$         &  $5\  10^{-2}$\\
 16   &  $1.3\  10^{8}$ 	&  $4\  10^{-3}$ & 47   &  $1.0\ 10^{6}$  	    &  $2\  10^{-2}$  & 78 &  $4.0 \  10^{7}$         &  $5\  10^{-2}$\\
 17   &  $1.4\  10^{9}$     	&  $4\  10^{-3}$ & 48   &  $3.2\  10^{6}$        &  $2\  10^{-2}$ &  79  &  $5.5 \  10^{7}$         & $5\  10^{-2}$\\
 18   &  $2.5\  10^{9}$     	&  $4\  10^{-3}$ & 49   &  $5.0\  10^{6}$        &  $2\  10^{-2}$ & 80   &  $1.0 \  10^{7}$         &  $5\  10^{-2}$ \\
 19   &  $4.3\  10^{9}$    	&  $4\  10^{-3}$ &  50   &  $6.6\  10^{6}$       &  $2\  10^{-2}$ & 81   &  $1.6 \  10^{8}$         & $5\  10^{-2}$\\
 20   &  $6.3\  10^{9}$    	&   $4\  10^{-3}$ & 51   &  $8.7 \  10^{6}$      &  $2\  10^{-2}$ & 82   &  $2.9\  10^{8}$         &  $5\  10^{-2}$ \\
 21   &  $7.5\  10^{9}$     	&   $4\  10^{-3}$ & 52   &  $1.0\ 10^{7}$        &  $2\  10^{-2}$ &  83   &  $5.1\  10^{8}$        & $5\  10^{-2}$ \\
 22   &  $1.0\ 10^{10}$    	&  $4\  10^{-3}$ & 53   &  $1.5\  10^{7}$        &  $2\  10^{-2}$ &  84   &  $ 9.0\  10^{8}$       &  $5\  10^{-2}$\\
 23   &  $1.3\  10^{10}$    	&  $4\  10^{-3}$ &  54   &  $2.5 \  10^{7}$      &  $2\  10^{-2}$ & 85    &  $1.3\  10^{8}$        &  $5\  10^{-2}$\\
 24   &  $1.0\ 10^{6}$        & $8\  10^{-3}$ 	&  55   &  $4.0 \  10^{7}$       &  $2\  10^{-2}$ &  86  &  $1.4\  10^{9}$        &  $5\  10^{-2}$\\
 25   &  $3.2\  10^{6}$       &  $8\  10^{-3}$ & 56   &  $5.5 \  10^{7}$        & $2\  10^{-2}$ & 87   &  $2.5\  10^{9}$         &  $5\  10^{-2}$\\
 26   &  $5.0\  10^{6}$       &  $8\  10^{-3}$ & 57   &  $1.0 \  10^{7}$        & $2\  10^{-2}$ & 88   &  $4.3\  10^{9}$        &  $5\  10^{-2}$ \\
 27   &  $6.6\  10^{6}$       &  $8\  10^{-3}$ &  58   &  $1.6 \  10^{8}$       & $2\  10^{-2}$ & 89   &  $6.3\  10^{9}$        &  $5\  10^{-2}$ \\
 28   &  $8.7 \  10^{6}$      &  $8\  10^{-3}$ & 59   &  $2.9\  10^{8}$        & $2\  10^{-2}$ & 90    &  $7.5\  10^{9}$         &  $5\  10^{-2}$\\
 29   &  $1.0\ 10^{7}$        &  $8\  10^{-3}$ & 60   &  $5.1\  10^{8}$        &  $2\  10^{-2}$ &  91  &  $1.0\ 10^{10}$       &   $5\  10^{-2}$ \\
 30   &  $1.5\  10^{7}$       &   $8\  10^{-3}$ &  61   &  $ 9.0\  10^{8}$     & $2\  10^{-2}$  &92    &  $1.3\  10^{10}$      &   $5\  10^{-2}$ \\
 31   &  $2.5\  10^{7}$       &  $8\  10^{-3}$  & 62   &  $1.3\  10^{8}$       & $2\  10^{-2}$ & & &  \\
\hline                                   
\end{tabular}
\label{table1}
\caption{List of stellar populations used for the evaluation of template spectra with age and metallicity $Z$.}
\end{table}

\begin{figure*}[htp!]
\centering
\includegraphics[width=12cm]{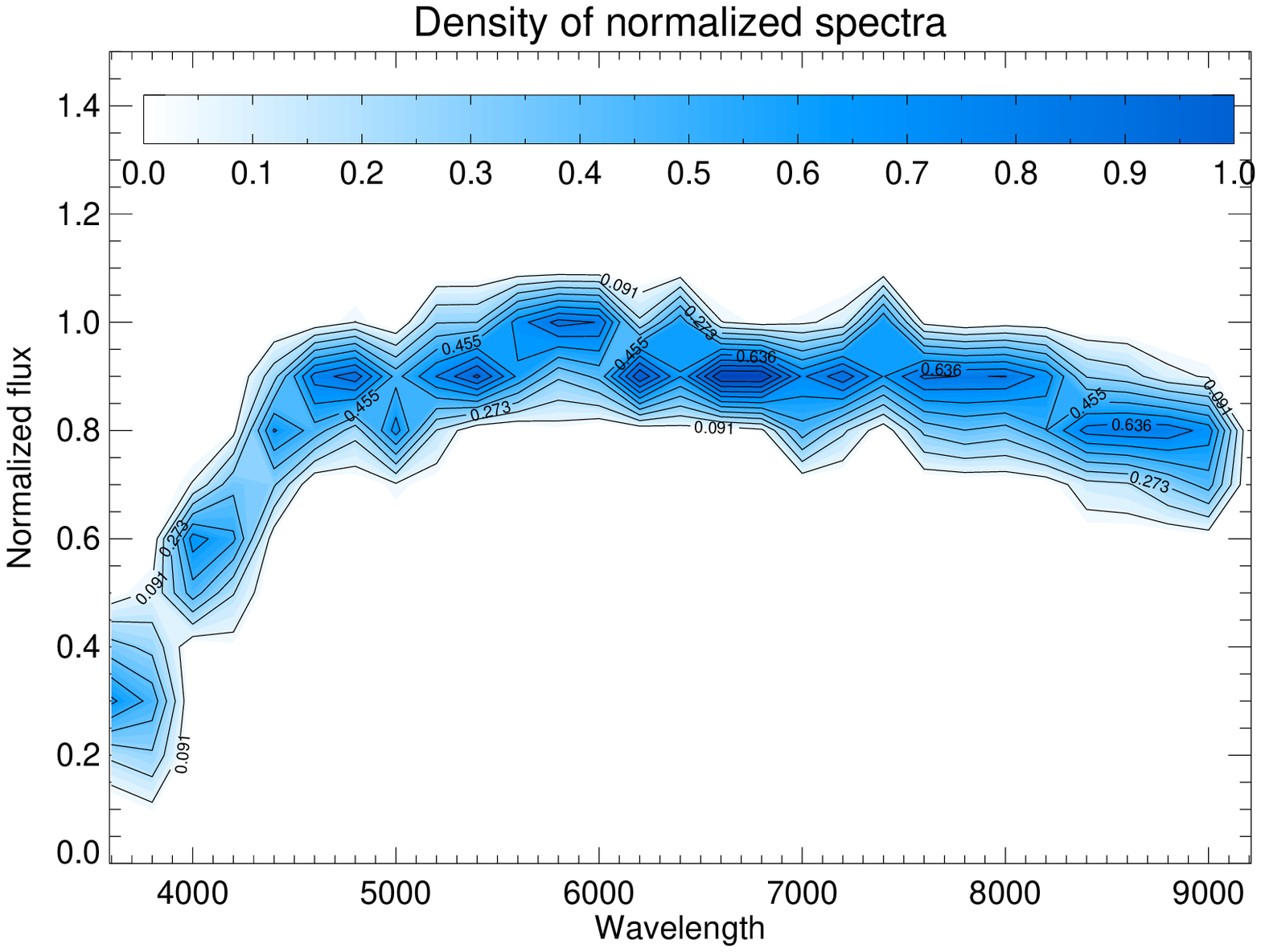}
\caption{Density of normalized spectra of the normal galaxies with isodensity contours.}
\label{denS2GdNGs}
\end{figure*}

\begin{figure*}[htp!]
\centering
\includegraphics[width=12cm]{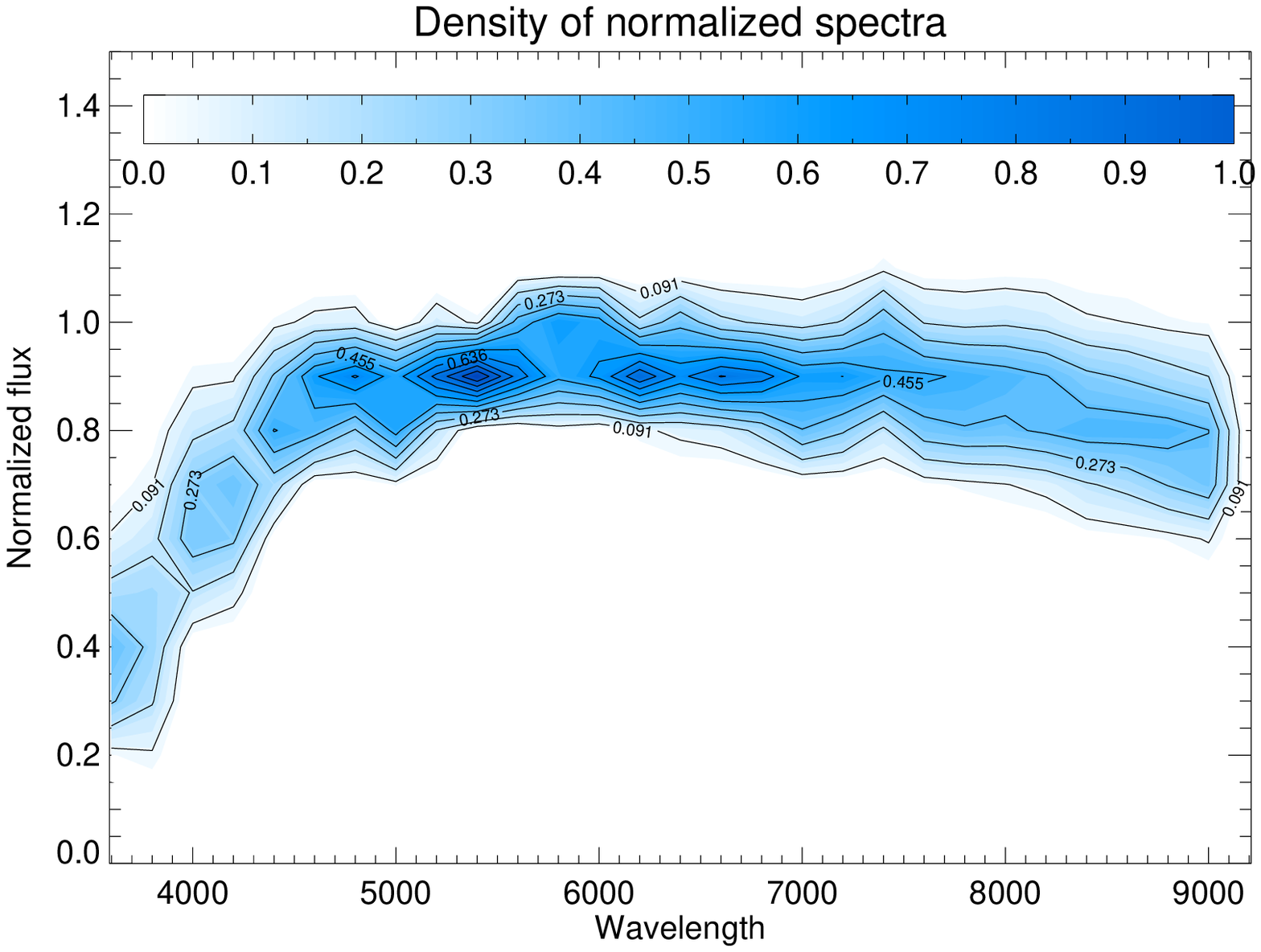}
\caption{Density of normalized spectra of the S2G galaxies with isodensity contours.}
\label{denS2GdS2G}
\end{figure*}

\begin{figure*}[htp!]
\centering
\includegraphics[width=12cm]{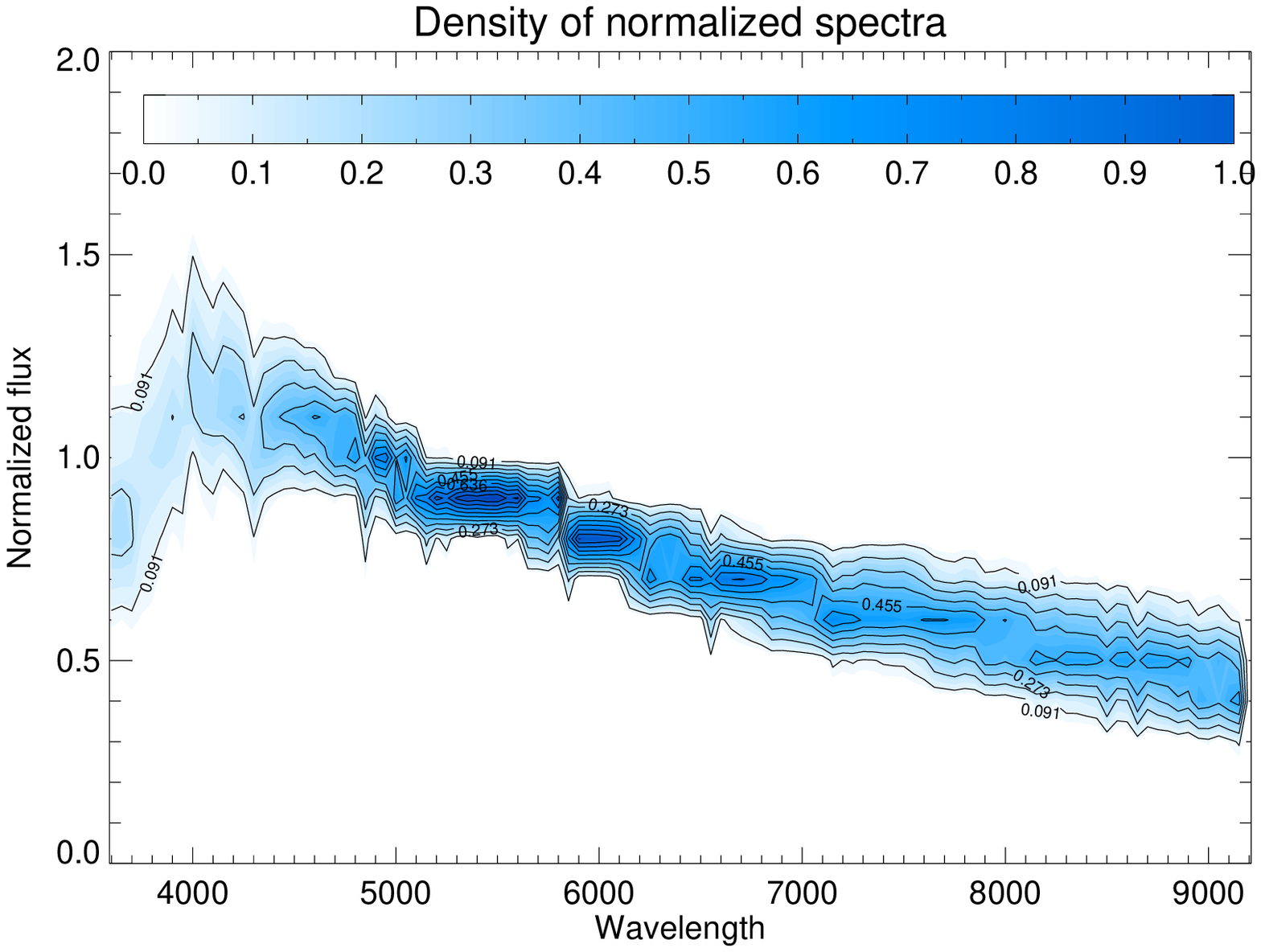}
\caption{Density of normalized spectra of the SFGs galaxies with isodensity contours.}
\label{denS2GdSFGs}
\end{figure*}

The template spectra of all galaxies of the three samples have been normalized to the spectral flux measured at $\lambda = 5500$\AA, since this
region of the spectrum is devoid of emission and absorption lines, thus being a good approximation of the continuum component 
of the spectrum. The general features of the normalized template spectra for each of the three samples of galaxies have 
been determined by overplotting the spectra and determining the bi-dimensional densities in the $(\lambda,\  F_{\lambda})$ plane, 
with the isondensity contours enclosing the regions of the plane where it is more likely to find the spectra (the region enclosed in the
outer isodensity curve containing $99\%$ of the spectra). The 2-dimensional 
densities for NGs, SFG galaxies and S2G are shown in figures \ref{denS2GdNGs}, \ref{denS2GdS2G} and \ref{denS2GdSFGs} respectively. 

\begin{figure*}[htp!]
\centering
\includegraphics[width=12cm]{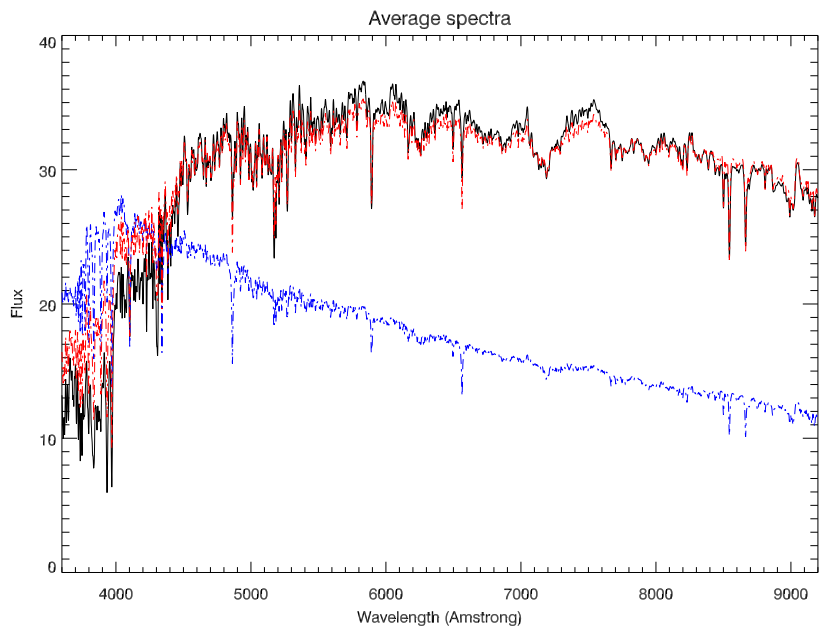}
\caption{Average spectra for NGs (black curve), S2G galaxies (red curve) and SFGs galaxies (blue curve), flux unit is   $10^{-17} erg\ \ cm^{-2} s^{-1} \AA^{-1}$}
\label{avgspectra}
\end{figure*}

\begin{figure*}[htp!]
\includegraphics[width=12cm]{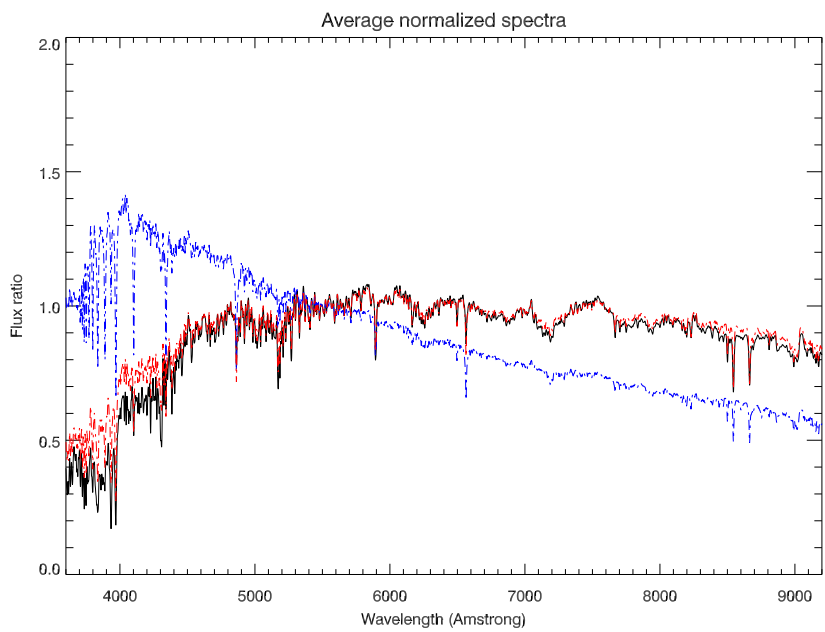}
\caption{Normalized average spectra for NGs (black curve), S2G galaxies (red curve) and SFGs galaxies (blue curve)}
\label{avgspectranorm}
\end{figure*}

\begin{figure*}[htp!]
\centering
\includegraphics[width=12cm]{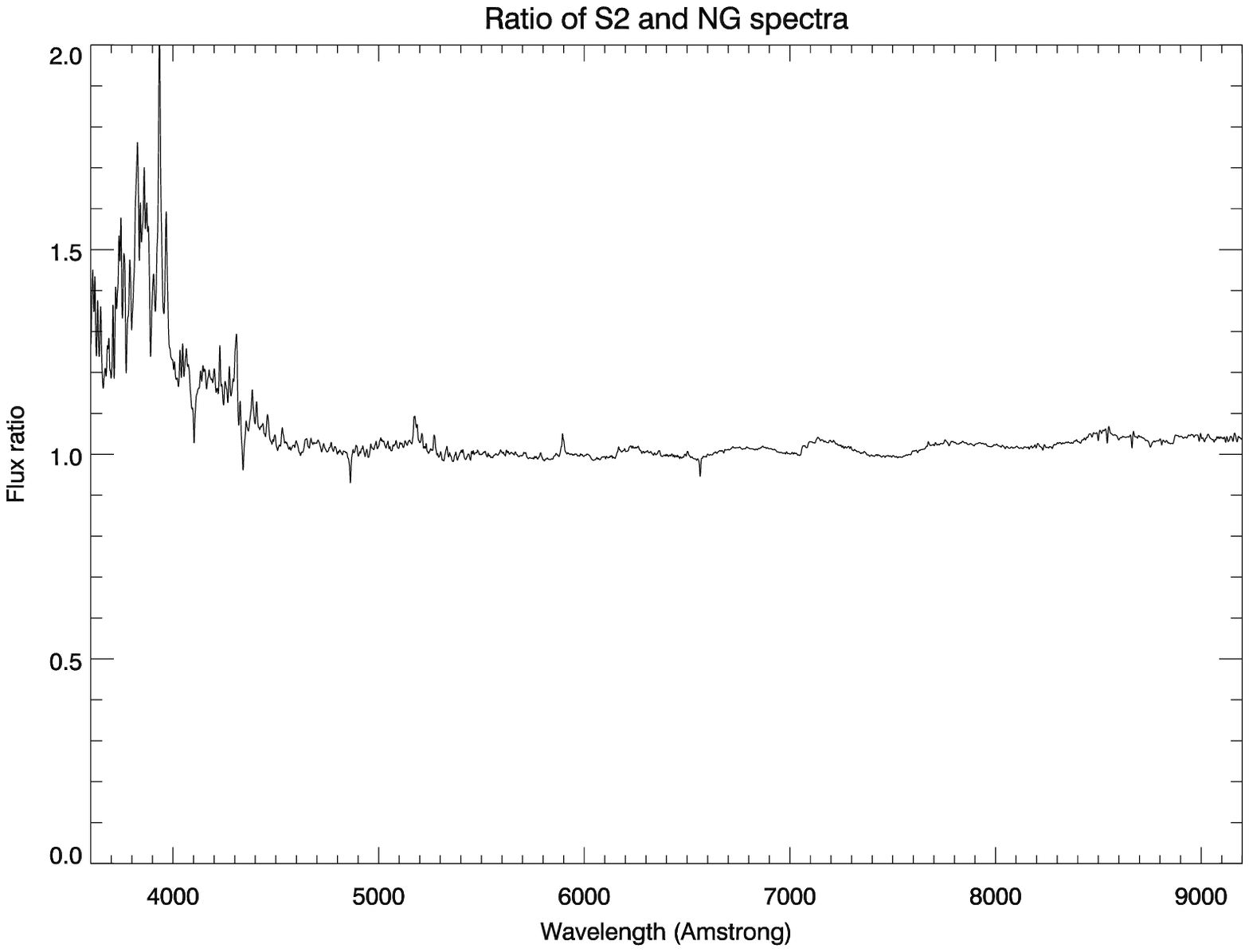}
\caption{Ratio of the normalized average spectra of NGs and S2G galaxies.}
\label{ratiospectra}
\end{figure*}

In figure \ref{avgspectra}, the average spectra derived by the distributions of spectra of the three classes of galaxies
are shown together, while the comparison of the normalized average spectra is shown in figure \ref{avgspectranorm} (in both figures,
blue is used for SFG, red for S2G galaxies and black for NGs). While the average spectrum of SFGs galaxies is high in the 
region of shorter wavelengths, the behaviour of the spectra of S2G and NGs galaxies is very similar. The ratio of the average normalized 
spectra of S2G galaxies and NGs is shown in figure \ref{ratiospectra} in order to highlight the differences between their shapes. The average 
spectrum of S2G galaxies appears to be systematically higher than the  average spectrum of the NGs at wavelengths smaller than $\lambda = 5500 \AA$, thus indicating the 
presence of a small fraction of young stellar populations in the nuclear regions of galaxies that harbor a Seyfert 2 nucleus which are not 
found in the surrounding of the nuclei of no-emission lines galaxies.  

\section{Conclusions}
\label{conclusions}

The presence of a strong blue continuum and the absence of any 4000\AA \ \ break in the spectra of the analyzed SFG are the clear indication of the presence in their circumnuclear regions of an excess of young hot blue stars compared with NG and S2G. The ultraviolet radiation emitted by hot O-B stars ionizes  all heavy elements in their atmospheres and cancels the apparently continuum absorption at wavelengths $\lambda \leq$ 4500\AA, produced by their crowded excitation levels. The features of the stellar continuum component of the circumnuclear spectra of SFG are in good agreement with the results obtained from their emission lines, confirming that strong star formation processes are occurring. On the contrary, the spectra of the circumnuclear regions of S2G and NG indicate quite similar conditions: a dominating stellar component of type F (like usual in normal galaxies) and a 4000\AA \ \ break in NG deeper than in S2G of a factor $\sim$1.5. This indicates that the presence of an excess of A stars in the circumnuclear regions of S2G compared to NG is quite likely, suggesting that there have been in the recent past of S2G  slightly more star formation events than in NG. {\bf In conclusion, a preliminary analysis of our data indicates that Seyfert 2 nuclei occurr preferentially in environments where recent circumnuclear star formation processes were active, even if there is no compelling evidence that would connect present or past star formation activity to the nature of the central engine. The two phenomena, SF and AGN activity, may therefore be not causally connected to each other or, if so, through indirect or/and not clear physical mechanisms. This result suggests that the two phenomena have quite likely a common origin and that they are timely connected but neither necessarily coeval nor causally connected.}



\end{document}